\documentclass[journal=jctcce,manuscript=article,hyphens,maxauthors=60]{achemso}
\usepackage[T1]{fontenc}

\usepackage[version=3]{mhchem} % Forula subscripts using \ce{}
\usepackage{multirow}
\usepackage[dvipsnames,table,xcdraw]{xcolor}
\usepackage{booktabs}  % For professional table formatting
\usepackage{array}      % For better column control
\usepackage{amsmath}    % For mathematical formatting
\usepackage{float} % in the preamble
\usepackage{colortbl}
\usepackage{braket}
\usepackage{longtable}
\usepackage{rotating}
\usepackage{graphicx}
\usepackage{threeparttable}
\usepackage{subcaption}
\usepackage{bm}
\usepackage{nicefrac}
\usepackage{acro}
\usepackage[colorinlistoftodos]{todonotes}

\newcommand{\Lower}[1]{\smash{\lower 1.5ex \hbox{#1}}}

\usepackage{soul}

\DeclareAcronym{pybest}{
  short= PyBEST,
  long= Pythonic Black-box Electronic Structure Tool,
}
\DeclareAcronym{hf}{
  short= HF,
  long= Hartree-Fock,
}
\DeclareAcronym{cc}{
  short= CC,
  long= Coupled Cluster,
}
\DeclareAcronym{pccd}{
  short= pCCD,
  long= pair Coupled Cluster Doubles,
}
\DeclareAcronym{ccsd}{
  short= CCSD,
  long= \ac{cc} Singles Doubles,
}
\DeclareAcronym{ccd}{
  short= CCD,
  long= \ac{cc} Doubles,
}
\DeclareAcronym{eom}{
  short= EOM,
  long= Equation of Motion,
}
\DeclareAcronym{pccds}{
  short= EOM-pCCD+S,
  long= \ac{pccd} with Singles correction
}
\DeclareAcronym{ct}{
  short= CT,
  long= charge transfer
}
\DeclareAcronym{dct}{
  short= dCT,
  long= directed charge transfer
}
\DeclareAcronym{daisy}{
  short= DAISpY,
  long= Domain Assignment and Interface Solution in pYthon
}
\DeclareAcronym{gui}{
  short= GUI,
  long= Graphical User Interface
}
\DeclareAcronym{mo}{
  short= MO,
  long= molecular orbital
}
\DeclareAcronym{ao}{
  short= AO,
  long= atomic orbital
}
\DeclareAcronym{ci}{
  short= CI,
  long= Configuration Interaction,
}

\author{Lena Szczuczko}
\affiliation{Institute of Physics, Faculty of Physics, Astronomy, and Informatics, Nicolaus Copernicus University in Toruń, Grudziądzka 5, 87-100 Toru\'{n}, Poland.}
\author{Julia Szczuczko}
\affiliation{Institute of Physics, Faculty of Physics, Astronomy, and Informatics, Nicolaus Copernicus University in Toruń, Grudziądzka 5, 87-100 Toru\'{n}, Poland.}
\author{Marta Ga\l{}y\'{n}ska}
\affiliation{Faculty of Chemistry, Nicolaus Copernicus University in Toru\'{n}, Gagarina 7, 87-100 Toru\'{n}, Poland}
\email{marta.galynska@umk.pl}
\author{Katharina Boguslawski}
\email{k.boguslawski@umk.pl}
\affiliation{Institute of Physics, Faculty of Physics, Astronomy, and Informatics, Nicolaus Copernicus University in Toruń, Grudziądzka 5, 87-100 Toru\'{n}, Poland.}

\title[]
{A {Flexible,} Automated{,} and Basis-Set {Insensitive} Domain-Based Charge-Transfer Decomposition for Correlated Wavefunctions and its Application to Inter- and Intramolecular Cases}
\abbreviations{CC, EOM-CC, HF, oo-pCCD}

\begin{document}

%%%%%%%%%%%%%%%%%%%%%%%%%%%%%%%%%%%%%%%%%%%%%%%%%%%%%%%%%%%%%%%%%%%%%
%% The "tocentry" environment can be used to create an entry for the
%% graphical table of contents. It is given here as some journals
%% require that it is printed as part of the abstract page. It will
%% be automatically moved as appropriate.
%%%%%%%%%%%%%%%%%%%%%%%%%%%%%%%%%%%%%%%%%%%%%%%%%%%%%%%%%%%%%%%%%%%%%
\begin{tocentry}

%\pawel{Please add a TOC according to guideline below (see the commented link)}
%https://pubsapp.acs.org/paragonplus/submission/toc_abstract_graphics_guidelines.pdf?_gl=1*5qlkne*_ga*Njg2NDYzNzA0LjE3MzYyODQ5NTA.*_ga_XP5JV6H8Q6*czE3NTE5ODk4MzQkbzQ1JGcwJHQxNzUxOTg5ODU0JGo0MCRsMCRoMA..*_ga_3YE6YD0SWD*czE3NTE5ODk4MzQkbzEkZzAkdDE3NTE5ODk4NTQkajQwJGwwJGgw&

% \includegraphics[width=8.25cm,height=4.45cm]{toc.png}
\centering
\includegraphics[height=4.45cm]{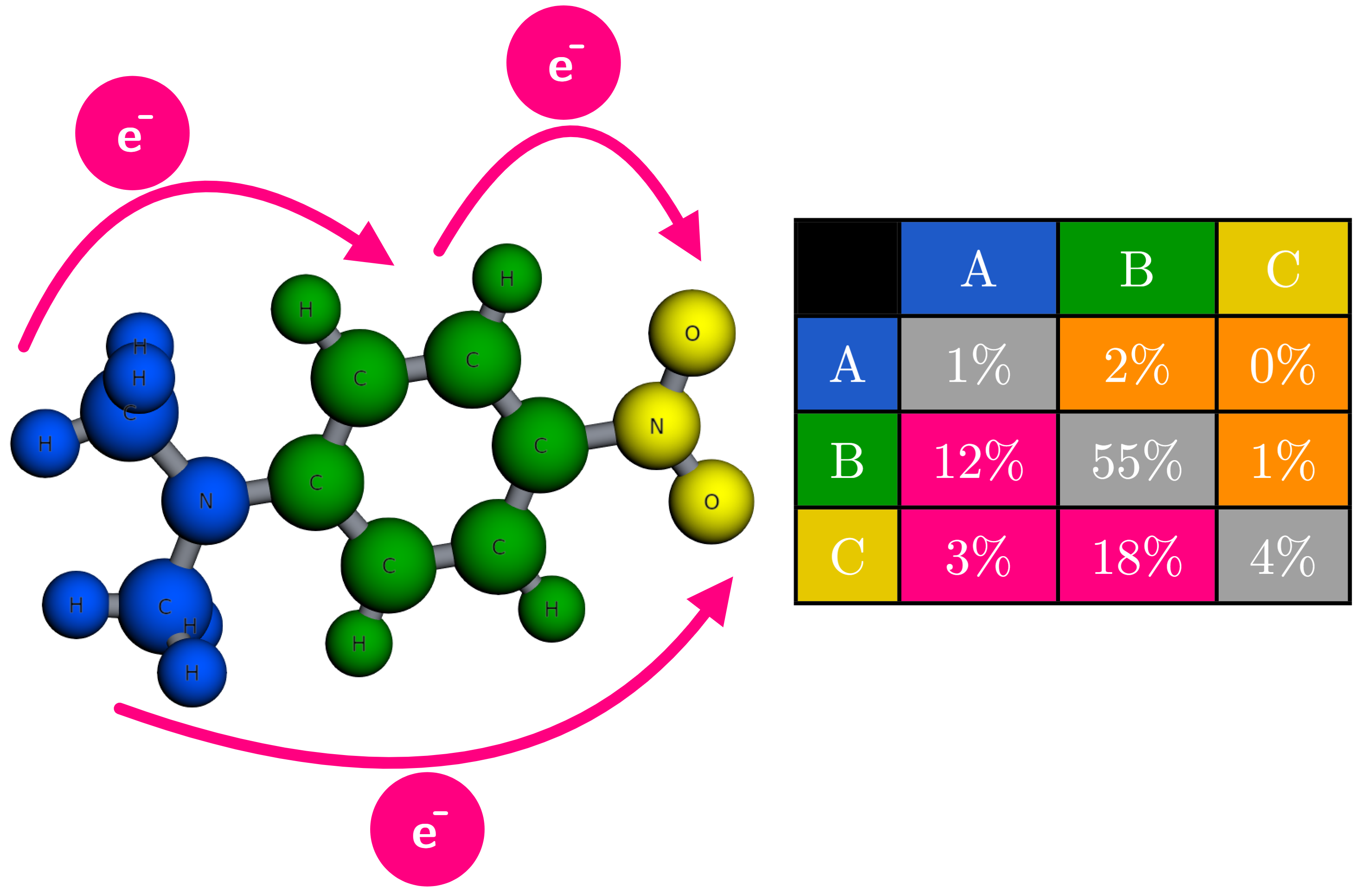}

\end{tocentry}

%%%%%%%%%%%%%%%%%%%%%%%%%%%%%%%%%%%%%%%%%%%%%%%%%%%%%%%%%%%%%%%%%%%%%
%% The abstract environment will automatically gobble the contents
%% if an abstract is not used by the target journal.
%%%%%%%%%%%%%%%%%%%%%%%%%%%%%%%%%%%%%%%%%%%%%%%%%%%%%%%%%%%%%%%%%%%%%
\begin{abstract}
We present a {flexible,} automated, and basis-set {insensitive} domain-based charge-transfer (CT) decomposition framework that can be combined with any CI-type excited-state wavefunction.
Our approach {is not based on excited-state densities and} allows the excited-state character to be dissected into local and domain-based CT excitations and measures the individual contributions to each excited state.
To guarantee a broad applicability, we introduce two domain-accumulation strategies to translate hole-particle substitutions to domain-domain excitations: a strict domain partitioning and a weighted approach suitable for small molecules and a large number of domains.
The performance of both schemes is assessed for inter- and intramolecular CT excitations and various basis sets using EOM-CCSD and its simplified counterpart EOM-pCCD+S.
Most importantly, the CT character is, to a large extent, basis-set independent, and both domain-accumulation schemes give consistent results.
Overall, our framework provides a robust CT analysis and a domain resolution of the excitation character for a variety of computational setups and excited-state models.
\end{abstract}

%%%%%%%%%%%%%%%%%%%%%%%%%%%%%%%%%%%%%%%%%%%%%%%%%%%%%%%%%%%%
%% Start the main part of the manuscript here.
%%%%%%%%%%%%%%%%%%%%%%%%%%%%%%%% introduction %%%%%%%%%%%%%%%%%%%%%%%%%%%%%%%

%%%%%%%%%%%%%%%%%%%%%%%%%%%%%%%%%%%%%%%%%%%%%%%%%%%%%%%%%%%%
%%%%%%%%%%%%%%%%%%%%%%%%%%%%%%%%%%%%%%%%%%%%%%%%%%%%%%%%%%%%
%%%%%%%%%%%%%%%%%%%%%%%%%%%%%%%%%%%%%%%%%%%%%%%%%%%%%%%%%%%%
% \section{Introduction} \label{sec:Introduction}
%%%%%%%%%%%%%%%%%%%%%%%%%%%%%%%%%%%%%%%%%%%%%%%%%%%%%%%%%%%%
%%%%%%%%%%%%%%%%%%%%%%%%%%%%%%%%%%%%%%%%%%%%%%%%%%%%%%%%%%%%
%%%%%%%%%%%%%%%%%%%%%%%%%%%%%%%%%%%%%%%%%%%%%%%%%%%%%%%%%%%%
\Ac{ct}\cite{ct_review_pccp} plays a central role in key biological processes, such as photosynthesis\cite{photosynthesis} and cellular respiration\cite{respiratory, mitochondria}, as well as in the operation of electronic devices, including photovoltaics, organic light-emitting diodes (OLEDs), and chemical sensors.~\cite{risko2011,p-block1-elements,joule-15-precent-osc,p-block2-elements, Isoelectronic1, p-block3-elements, optoelectronics}
\ac{ct} may take place within a single molecular entity (intramolecular \ac{ct}) or between distinct molecular species or across interfaces (intermolecular \ac{ct}). 
Depending on the mechanism, it can be photoinduced, arising from electronically excited states or between different conformational states on the same potential energy surface. 
Despite this diversity, the \ac{ct} process governs the redistribution of electronic density between molecular fragments and is driven by the presence of electron-rich (donor) and electron-deficient (acceptor) fragments. 
A detailed understanding of the spatial distribution of charge across those molecular domains is of particular interest. 
Especially in terms of electronic excitations, it is essential to characterize the nature of the excited state, particularly whether an excitation is local or involves \ac{ct} between fragments.
However, the description of electronic excitations depends critically on the theoretical framework employed, and the methodology used to analyze excited states must be chosen accordingly.

Understanding the spatial character of electronic excitations, specifically, whether they are local or involve \ac{ct}, constitutes an important aspect of excited-state chemistry.
Excited-state analysis methods can be broadly categorized by how electronic excitations are represented.
In linear-response approaches, such as time-dependent density functional theory (TD-DFT)\cite{td-dft} or linear response coupled cluster (LR-CC)\cite{monkhorst_1977, dalgaard_monkhorst_1983}, excited states are treated as perturbations of the ground state.
In this formalism, the transition density matrix (TDM), which couples the ground and excited states, arises naturally as a central quantity.~\cite{tdm_ct_jctc_2014}
Together with density difference, they provide a compact representation of electronic excitations in terms of electron–hole transitions and enable the construction of several descriptors such as natural transition orbitals, attachment/detachment analysis, fragment-based analysis, exciton size, average position of hole and particle, charge transfer distance, or hole/electron delocalization, to name a few.\cite{nto_jcp_2003, plasser_descriptors_jctc_2012, plasser_jcp_2014, spatial_extend_ct_jctc_2011, ct_metric_jctc_2013, overlap_exciton_jcp_2008, ghost_hunter_index_jcc_2017, dm_analysis_cr_2002, emd_ct_jctc_2023, tdm_ct_jctc_2014}
This type of analysis is available in dedicated tools such as Theo{DORE}\cite{plasser2020theodore}, Multiwfn\cite{multiwfn_jcp_2024}, VALET\cite{valet_2021}, or implemented into quantum chemistry packages such as Gaussian\cite{g16}, Orca\cite{ORCA, ORCA5}, Q-Chem\cite{qchem_2015}, OpenMolcas\cite{open_molcas}, or Veloxchem\cite{veloxchem_wires_2020, echem_jce_2023}. 
In contrast, \ac{eom} for excitation energies~\cite{eom-cc-rowe-rmp-1968, eom-ccsd-bartlett-jcp-1993, bartlett-eom-cc-wires-2012} and related formalisms~\cite{sacci-naktsuji-cpl-1991,ip-eom-cc-ijqc-1992,ip-eom-cc-ijqc-1993,ea-eom-cc-marcel-jcp-1995,sf-eom-ccsd-jcp-2008} describe excited states as explicit many-electron states built from the ground-state reference via excitation operators.
In these approaches, the excited-state wavefunction is represented in terms of \ac{ci} expansion coefficients, which directly encode the contributions of individual orbital transitions.
While transition density matrices can also be derived within the EOM framework, their construction requires additional effort, as both left and right eigenstates must be computed and combined.\cite{eom-ccsd-bartlett-jcp-1993}
Therefore, for this particular case, an alternative approach to excited-state analysis is desirable{, specifically, one that does not rely on excited-state densities}.

While excited states are rigorously defined through wavefunction expansions, interpreting \textit{where} electron density moves remains non-trivial.
Traditional approaches, such as orbital inspection or density difference plots or natural transition orbitals, often provide only qualitative insight and may depend strongly on the chosen representation and individual interpretation.
Domain-based approaches address this limitation by introducing an explicit spatial partitioning of the system. 
Recently, we introduced a domain-based \ac{ct} analysis for electronically excited states computed with the simple EOM \ac{pccds} method~\cite{ap1rog-piotrus-jctc-2013,oo-ap1rog-prb-2014,tamar-pccd-jcp-2014,eom-pccd-jcp-2016,eom-pccd-erratum-jcp-2017,pccp-geminal-review-pccp-2022} and applied it to donor–bridge–acceptor dyes, designed as prototypical systems for dye-sensitized solar cell applications.~\cite{pccd-perspective-jpcl-2023, lena-ct-jctc-2025, ram_jpca_2026}
{Our original approach was based on a manual, tedious, and time-consuming evaluation of excited-state contributions and restricted to single excitations, which severely limits its applicability.}
Here, we generalize this approach and present a {flexible,} automated scheme to dissect the character of excited states into local and inter-domain excitations (or \ac{ct}-type excitations) for any correlated \ac{ci}-type wavefunction expansion and different choices for domain accumulation.
{Importantly, the framework applies to any correlated CI‑type excited‑state method that does not require explicit evaluation of left eigenvectors (in the case of EOM-type methods) or excited‑state densities. 
Nonetheless, our methodology can be generalized, and the CT analysis can be performed using the right eigenvector only (as discussed in this work), the left eigenvector in addition to the right ones, or an approximate left eigenvector constructed from the right one.\mbox{~\cite{left-evec-eom-neese-jcp-2017}}}
We test our approach for both intra- and intermolecular systems and scrutinize its basis set dependence. 
Specifically, we probe our analysis for \ac{eom}-\ac{ccsd} and compare the directed domain-based \ac{ct} character to other \ac{eom}-\ac{ccsd} data present in the literature, for both inter- and intra-molecular \ac{ct} excitations.
{We should stress that the aim of this work is to provide an alternative to density-based partition schemes of excited states that can be applied for any CI‑type excited state wave function.}
{We further aim at an efficient evaluation of our CT matrix and the derived CT indices.}
{Thus, we exploit the right eigenvectors only.}
{Although being an approximation within the EOM-CC formalism, our numerical data support this shortcut.}
{Note, however, that the proposed methodology is not restricted to this approximation.}
%%%%%%%%%%%%%%%%%%%%%%%%%%%%%%%%%%%%%%%%%%%%%%%%%%%%%%%%%%%%%%%%%%%%
%%%%%%%%%%%%%%%%%%%%%%%%%%%%%%%%%%%%%%%%%%%%%%%%%%%%%%%%%%%%%%%%%%%%
%%%%%%%%%%%%%%%%%%%%%%%%%%%%% theory %%%%%%%%%%%%%%%%%%%%%%%%%%%%%%%
% \section{Theory}\label{sec:Theory}
%%%%%%%%%%%%%%%%%%%%%%%%%%%%%%%%%%%%%%%%%%%%%%%%%%%%%%%%%%%%%%%%%%%%
%%%%%%%%%%%%%%%%%%%%%%%%%%%%%%%%%%%%%%%%%%%%%%%%%%%%%%%%%%%%%%%%%%%%
%%%%%%%%%%%%%%%%%%%%%%%%%%%%%%%%%%%%%%%%%%%%%%%%%%%%%%%%%%%%%%%%%%%%

\begin{figure}[ht]
\centering
\includegraphics[width=0.4\textwidth]{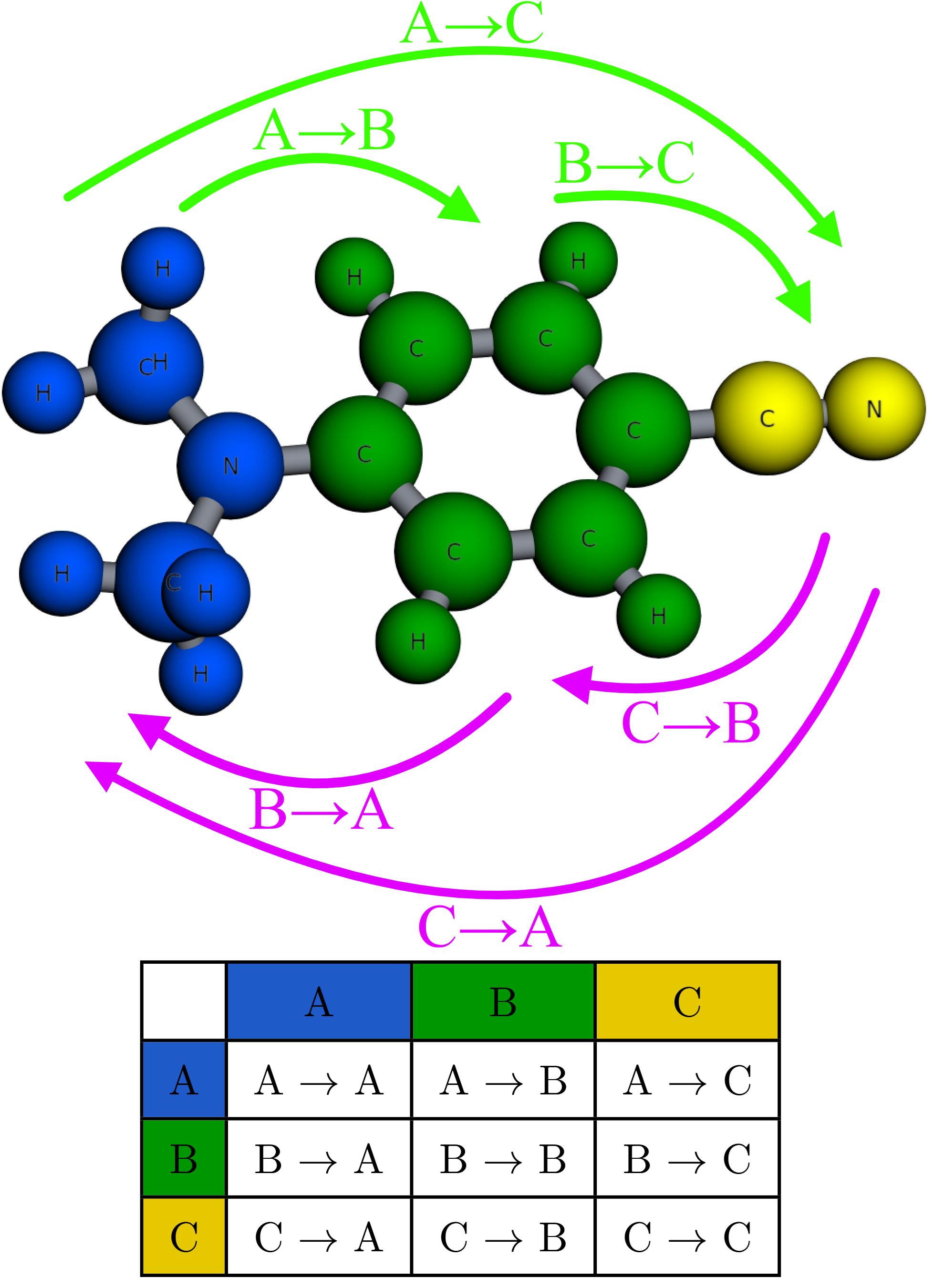}
\caption{
    Schematic representation of a three-domain \ac{ct} analysis across a donor--bridge--acceptor system.
    The corresponding domain-domain excitation matrix is shown below, comprising the individual initial--to--final domain excitations.
    The donor {(A, blue), bridge (B, green), and acceptor (C, yellow)} define forward (green arrows) and backward (magenta arrows) charge transfer pathways. 
    The net directed CT is obtained by summing all forward contributions {(A→B, B→C, A→C)} and subtracting the corresponding reverse processes {({A}←B, B←{C}, {A}←{C})}.
    }
\label{fig:1}
\end{figure}

%%%%%%%%%%%%%%%%%%%%%%%%%%%%%%%%%%%%%%%%%%%%%%%%%%%%%%%%%%%%%%%%%%%%
%%%%%%%%%%%%%%%%%%%%%%%%%%%%%%%%%%%%%%%%%%%%%%%%%%%%%%%%%%%%%%%%%%%%
% \subsection{A fully automated \ac{ct} decomposition}
%%%%%%%%%%%%%%%%%%%%%%%%%%%%%%%%%%%%%%%%%%%%%%%%%%%%%%%%%%%%%%%%%%%%
%%%%%%%%%%%%%%%%%%%%%%%%%%%%%%%%%%%%%%%%%%%%%%%%%%%%%%%%%%%%%%%%%%%%
In our {flexible,} automated, domain-based \ac{ct} framework, the molecule is divided into chemically relevant regions (domains), and excitations are analyzed in terms of electron transfer between these regions (see also Figure~\ref{fig:1} for a schematic representation).
Our framework is valid for any \ac{ci}-type excited-state wavefunction, where the excited state of interest $\ket{\Psi_k}$ is parameterized as
\begin{equation}\label{eq:exstate}
    \ket{\Psi_k} = \sum_{\mu = 0,S,D,\ldots} c_{\mu} \hat{\tau}_{\mu} \ket{\Psi_0},
\end{equation}
where the sum runs over all possible single, double, etc.~excitations and $\hat{\tau}_{\mu}$ represents an excitation operator (supplemented by the identity operator $\hat{\tau}_0$) that promotes electrons from the occupied to the virtual orbital space (related to some reference determinant), while $\ket{\Psi_0}$ encodes the ground-state wavefunction.
In the following, we will benchmark our domain-based CT framework for excited states modeled within the EOM-CCSD model.
Thus, we will restrict the above ansatz to single and double excitations and derive the CT measures within this excitation manifold.
However, our approach can be generalized to higher-order excitations.
Restricting the ansatz to single and double excitations, eq.~\eqref{eq:exstate} can be cast into
%%%%%%%%%%%%%%%%%%%%%%%%%%%%%%%%%%%%%%%%%%%%%%%%%%%%%%%%%%%%%%%%%%%%
\begin{equation}\label{eq:exstatesd}
    \ket{\Psi_k^{SD}} = \Big( 
        c_{0} \hat{\tau}_{0}
        + \sum_{i}^{\text{occ}} \sum_{a}^{\text{virt}} c_i^a \hat{\tau}_{ai}
        + \sum_{i,j}^{\text{occ}} \sum_{a,b}^{\text{virt}} c_{ij}^{ab} \hat{\tau}_{abij}
        \Big)
        \ket{\Psi_0},
\end{equation}
%%%%%%%%%%%%%%%%%%%%%%%%%%%%%%%%%%%%%%%%%%%%%%%%%%%%%%%%%%%%%%%%%%%%
where $\hat{\tau}_{ai}$ encodes single excitations (from an occupied orbital $i$ to a virtual orbital $a$) and $\hat{\tau}_{abij}$ creates a doubly-excited configuration.
We should note that we have dropped all prefactors in the above equation as they will depend on the chosen excited-state parameterization (spin-free vs.~spin-integrated formalisms).

%%%%%%%%%%%%%%%%%%%%%%%%%%%%%%%%%%%%%%%%%%%%%%%%%%%%%%%%%%%%%%%%%%%%
%%%%%%%%%%%%%%%%%%%%%%%%%%%%%%%%%%%%%%%%%%%%%%%%%%%%%%%%%%%%%%%%%%%%
% \subsection{The spin-free \ac{ccsd} and \ac{eom} ansatz}
%%%%%%%%%%%%%%%%%%%%%%%%%%%%%%%%%%%%%%%%%%%%%%%%%%%%%%%%%%%%%%%%%%%%
%%%%%%%%%%%%%%%%%%%%%%%%%%%%%%%%%%%%%%%%%%%%%%%%%%%%%%%%%%%%%%%%%%%%
In \ac{cc} theory, the correlated many-electron wavefunction is expressed by an exponential ansatz acting on a single-reference determinant $|\Phi_0\rangle$, usually a \ac{hf} determinant.
In this work, we consider two restrictions of the general \ac{cc} ansatz, namely, we restrict our analysis to a \ac{ccsd} ground-state reference function, and we focus on spin-free, singlet excitations solely.
The corresponding \ac{ccsd} ansatz,
%%%%%%%%%%%%%%%%%%%%%%%%%%%%%%%%%%%%%%%%%%%%%%%%%%%%%%%%%%%%%%%%%%%%
\begin{equation}
    |\Psi_{\rm CCSD}\rangle = e^{\hat{T}_1 + \hat{T}_2} |\Phi_0\rangle
\end{equation}
%%%%%%%%%%%%%%%%%%%%%%%%%%%%%%%%%%%%%%%%%%%%%%%%%%%%%%%%%%%%%%%%%%%%
then includes spin-free single ($\hat T_1$) and double ($\hat T_2$) excitation operators, which are defined through the singlet excitation operator $E^a_i = \hat a_a^\dagger \hat a_i + \hat a_{\bar a}^\dagger \hat a_{\bar i}$ as follows
%%%%%%%%%%%%%%%%%%%%%%%%%%%%%%%%%%%%%%%%%%%%%%%%%%%%%%%%%%%%%%%%%%%%
\begin{equation}
\hat{T}_1 = \sum_{i}^{\mathrm{occ}} \sum_{a}^{\mathrm{virt}} t_i^a E^a_i
\end{equation}
%%%%%%%%%%%%%%%%%%%%%%%%%%%%%%%%%%%%%%%%%%%%%%%%%%%%%%%%%%%%%%%%%%%%
and
%%%%%%%%%%%%%%%%%%%%%%%%%%%%%%%%%%%%%%%%%%%%%%%%%%%%%%%%%%%%%%%%%%%%
\begin{equation}
\hat{T}_2 = \frac{1}{2}
    \sum_{i,j}^{\mathrm{occ}}
    \sum_{a,b}^{\mathrm{virt}}
        t_{ij}^{ab}
        E^a_i E^b_j,
\end{equation}
%%%%%%%%%%%%%%%%%%%%%%%%%%%%%%%%%%%%%%%%%%%%%%%%%%%%%%%%%%%%%%%%%%%%
where $t_i^a$ and $ t_{ij}^{ab}$ denote single and double excitation amplitudes, respectively, and $\hat{a}_a^\dagger$ and $\hat{a}_i$ are the creation and annihilation operators for $\alpha$ ($a$) and $\beta$ ($\bar a)$ spin.
In this spin-free picture, the excitation operator $\hat \tau$ in eq.~\eqref{eq:exstatesd} can be identified as $\hat{\tau}_{ai} = E^a_i$ and $\hat{\tau}_{abij} = E^a_i E^b_j$, respectively.
Despite these restrictions, our analysis remains valid for the spin-integrated case.
Assuming spin-free excitations allows for a compressed representation of excited states, which reduces the overall storage of excitation amplitudes and thus speeds up the underlying analysis.

%%%%%%%%%%%%%%%%%%%%%%%%%%%%%%%%%%%%%%%%%%%%%%%%%%%%%%%%%%%%%%%%%%%%
%%%%%%%%%%%%%%%%%%%%%%%%%%%%%%%%%%%%%%%%%%%%%%%%%%%%%%%%%%%%%%%%%%%%

To resolve an excited state into domain contributions, we introduce a domain--domain excitation matrix, which comprises both local (or intra-domain) and inter-domain excitations (see Figure~\ref{fig:1} for a three-domain example).
Specifically, this domain--domain excitation matrix is constructed by accumulating CI contributions according to the domain character of the participating orbitals (that is, deduced from hole-particle excitations). 
For a given orbital $p$ and domain $D$, we define a scheme-dependent domain factor
\begin{equation}
\Omega_D(p)=
\begin{cases}
1 \ \text{if } D=\displaystyle\arg\max_{D'}\sum_{\mu\in D'}|C_{\mu p}|^2,\ \text{otherwise }0 & \text{(discrete or hard scheme)}\\[8pt]
\displaystyle\sum_{\mu\in D}|C_{\mu p}|^2 & \text{(weighted scheme)}
\end{cases}
\end{equation}
where $C_{\mu p}$ are the orthonormalized \ac{ao} expansion coefficients of molecular orbital $p$, and the sum runs over all atomic basis functions $\mu$ belonging to domain $D$.
Note that $C_{\mu p}$ are obtained from the non-orthonormal expansion coefficients through $S^{1/2}_{\mu \nu}\tilde C_{\nu p}$, where $S$ is the overlap matrix of the \ac{ao} basis.
{For the hard scheme, $\Omega_D(p)$ takes only the discrete values 0 or 1, indicating whether orbital $p$ is assigned to domain $D$ (1) or not (0). 
In contrast, in the weighted scheme $\Omega_D(p)$ assumes any value between 0 and 1 and reflects how strongly orbital $p$ is localized on domain $D$, as determined by the AO–MO expansion coefficients. 
This distinction clarifies that the hard scheme enforces a strict domain assignment, whereas the weighted scheme provides a continuous measure of domain contribution.}
In the context of atom-centered basis sets, a domain is, hence, defined as a collection of atoms (see also Figure~\ref{fig:1}).
Using this definition, the domain-based excitation matrix contribution for single excitations is given by
\begin{equation}\label{eq:domain-singles}
M_{D_hD_p}^{(S)}=
\sum_{i}^{\textrm{occ}} \sum_a^{\textrm{virt}}|c_i^a|^2\,\Omega_{D_h}(i)\,\Omega_{D_p}(a),
\end{equation}
where $i$ and $a$ denote occupied and virtual orbitals, respectively, and $c_i^a$ are the corresponding CI amplitudes of the excited state in question.
Note that $M_{D_hD_p}$ contains the $D_h \to D_p$ domain excitation element, where $D_h$ is the initial domain (hole states) and $D_p$ the final domain (particle states).
Double excitations are implemented within a one-electron channel concept, in which each excitation $(i,j)\rightarrow(a,b)$ is decomposed into two contributions, yielding (in the spin-free formulation)
\begin{equation}\label{eq:domain-doubles}
M_{D_hD_p}^{(D)}=
\frac{1}{2} \sum_{i,j}^{\textrm{occ}} \sum_{a,b}^{\textrm{virt}}|c_{ij}^{ab}|^2
\left[
\Omega_{D_h}(i)\,\Omega_{D_p}(a)
+
\Omega_{D_h}(j)\,\Omega_{D_p}(b)
\right],
\end{equation}
where $c_{ij}^{ab}$ denote double-excitation amplitudes and $(i,j)$ and $(a,b)$ label occupied and virtual orbitals involved in the excitation.
We should stress that{, in the spin-free case,} eq.~\eqref{eq:domain-doubles} yields eq.~\eqref{eq:domain-singles} for the case of electron-pair excitations where $i=j$ and $a=b$.
{Note that, in the spin-free case, electron-pair excitations are composite single excitations which are given by the $c_{ii}^{aa}$ excited-state vector contribution.}
{We emphasize once more that all expressions and operator relations used in this work are formulated within the spin‑free framework.}

The resulting domain-based excitation matrix provides a compact representation of charge redistribution between domains.
Diagonal elements correspond to local excitations, while off-diagonal elements quantify inter-domain charge transfer.
As illustrated in Figure~\ref{fig:1}, these contributions can be interpreted as directed \ac{ct} pathways between domains.
We should note that the sum over all domain-based excitation matrix elements is the same for both scheme-dependent domain factors.
The outlined framework provides a general and method-independent approach to analyze the \ac{ct} character in excited states represented by CI-type wavefunctions. 
Its practical realization is implemented in the \ac{daisy} package,~\cite{jj-daispy-inprep-2026,daispy-gitlab} which automates domain construction, orbital assignment, and the accumulation of domain--domain excitation matrices for selected excited states. 

To quantify the directionality of the charge flow in, for instance, a donor–bridge–acceptor system (see Figure~\ref{fig:1}), we introduce a \ac{dct} measure.
This quantity accounts for all possible pairwise \ac{ct} contributions between the donor ({A}), bridge (B), and acceptor ({C}), while explicitly differentiating between forward (donor-to-bridge-to-acceptor) and backward (acceptor-to-bridge-to-donor) flow.
Positive contributions correspond to \ac{ct} along the forward direction, whereas negative contributions capture \ac{ct} in the opposite direction.
The resulting expression therefore provides a compact measure of the overall \ac{ct} across the system.
For a collection of $q$ domains \{$D_q$\}, the \ac{dct} from $D_1$ to $D_2$ to $D_3$ etc.~can be evaluated from
%%%%%%%%%%%%%%%%%%%%%%%%%%%%%%%%%%%%%%%%%%%%%%%%%%%%%%%%%%%%%%%%%%%%
\begin{equation}\label{eq:generic_dct}
\mathrm{dCT} = \sum_{D_h, D_p}^{D_1 \to \ldots \to D_q}
    \Big[
        D_h \to D_p - D_h \leftarrow D_p
    \Big]
\end{equation}
%%%%%%%%%%%%%%%%%%%%%%%%%%%%%%%%%%%%%%%%%%%%%%%%%%%%%%%%%%%%%%%%%%%%
For the donor--bridge--acceptor pathway described above and visualized in Figure~\ref{fig:1}, we explicitly obtain
%%%%%%%%%%%%%%%%%%%%%%%%%%%%%%%%%%%%%%%%%%%%%%%%%%%%%%%%%%%%%%%%%%%%
\begin{align}\label{eq:dct}
\mathrm{dCT}({A} \to B \to {C}) =\, 
    &\left[({A} \to B) + (B \to {C}) + ({A} \to {C})\right] \nonumber \\
    &- \left[({A} \leftarrow B) + (B \leftarrow {C}) + ({A} \leftarrow {C})\right].
\end{align}
%%%%%%%%%%%%%%%%%%%%%%%%%%%%%%%%%%%%%%%%%%%%%%%%%%%%%%%%%%%%%%%%%%%%
{This expression corresponds to taking the difference between the upper and lower triangular parts of the \mbox{\ac{ct}} matrix, excluding the diagonal.}
{Moreover, the donor, acceptor, and bridge labels used throughout this work are not assigned a priori. 
Instead, the roles of individual domains are determined a posteriori from the computed \mbox{\ac{dct}} flow. 
In other words, the domain classification emerges naturally from the analysis itself and the physico-chemical process under consideration, rather than being imposed beforehand, ensuring that the interpretation of charge redistribution is fully consistent with the underlying data.}

%%%%%%%%%%%%%%%%%%%%%%%%%%%%%%%%%%%%%%%%%%%%%%%%%%%%%%%%%%%%%%%%%%%%
%%%%%%%%%%%%%%%%%%%%%%%%%%%%%%%%%%%%%%%%%%%%%%%%%%%%%%%%%%%%%%%%%%%%
% \subsection{Unambiguous domain assignment procedure}
%%%%%%%%%%%%%%%%%%%%%%%%%%%%%%%%%%%%%%%%%%%%%%%%%%%%%%%%%%%%%%%%%%%%
%%%%%%%%%%%%%%%%%%%%%%%%%%%%%%%%%%%%%%%%%%%%%%%%%%%%%%%%%%%%%%%%%%%%
Finally, to obtain unambiguous domain-to-domain excitations, it is essential that the automated domain decomposition can be performed unequivocally, where hole or particle indices are uniquely associated with domain indices.
If our domain decomposition is performed on delocalized canonical orbitals, the corresponding domain accumulation might yield smeared-out weighting factors, deforming a domain-based \ac{ct} analysis.
For that purpose, all wavefunction optimizations are performed within a localized \ac{mo} basis.
In principle, we could employ any localized \acp{mo}, such as Pipek--Mezey~\cite{pipek_localizatoin_jcp_1989,pipek_localization_jctc_2013} ones or others.~\cite{boys_localization_rmp_1960,ruedenberg_localization_rmp_1963}
However, to obtain strongly localized virtual orbitals, we exploit \ac{pccd}-optimized~\cite{ap1rog-piotrus-jctc-2013, oo-ap1rog-prb-2014, tamar-pccd-jcp-2014, pccp-geminal-review-pccp-2022} \acp{mo}.
Unlike the standard \ac{ccd} approach, which includes all possible double excitations, \ac{pccd} employs a simplified ansatz that selectively accounts only for paired electron excitations. 
The \ac{pccd} wavefunction is formulated using an exponential ansatz,~\cite{ap1rog-non-variational-orbital-optimization-jctc-2014, tamar-pccd-jcp-2014, pccp-geminal-review-pccp-2022}
%%%%%%%%%%%%%%%%%%%%%%%%%%%%%%%%%%%%%%%%%%%%%%%%%%%%%%%%%%%%%%%%%%%%
\begin{equation}
    \ket{\Psi_{\rm pCCD}} = 
        e^{\sum_{i}^{\text{occ}} \sum_{a}^{\text{virt}} t_{i\bar i}^{a \bar a} \hat{a}_a^\dagger \hat{a}_{\bar{a}}^\dagger \hat{a}_{\bar{i}} \hat a_i}
        \ket{\Phi_0} 
        = e^{\hat{T}_2^{\text{pCCD}}} \ket{\Phi_0},
\end{equation}
%%%%%%%%%%%%%%%%%%%%%%%%%%%%%%%%%%%%%%%%%%%%%%%%%%%%%%%%%%%%%%%%%%%%
where the \acp{mo} used to construct the reference determinant are typically optimized using a variational orbital optimization procedure.~\cite{tamar-pccd-jcp-2014,ps2-ap1rog-jcp-2014,ap1rog-non-variational-orbital-optimization-jctc-2014}
The optimal set of orbitals results in a vanishing \ac{pccd} orbital gradient (in spatial orbitals),
%%%%%%%%%%%%%%%%%%%%%%%%%%%%%%%%%%%%%%%%%%%%%%%%%%%%%%%%%%%%%%%%%%%%
\begin{align}\label{eq:voo-grad}                                           
% \frac{\partial \mathcal{L}}{\partial \kappa_{pq}}\Big\vert_{\bm \kappa=0} = 
g_{pq}  &=
    \langle \Phi_0 | e^{-\hat{T}_2^{\text{pCCD}}} [(\hat E_{pq} - \hat E_{qp}),\hat{H}]e^{\hat{T}_2^{\text{pCCD}}} | \Phi_0 \rangle \nonumber \\
    &\phantom{=} + \sum_{i,a} \lambda_{i\bar i}^{a \bar a} \big( \langle \Phi_{i \bar{i}}^{a \bar{a}} | e^{-\hat{T}_2^{\text{pCCD}}} [(\hat E_{pq} - \hat E_{qp}), \hat{H} ]e^{\hat{T}_2^{\text{pCCD}}} | \Phi_0\rangle \big) \nonumber \\
    &= 0 \quad \forall p>q,
\end{align}
%%%%%%%%%%%%%%%%%%%%%%%%%%%%%%%%%%%%%%%%%%%%%%%%%%%%%%%%%%%%%%%%%%%%
where $\lambda_{i\bar i}^{a \bar a}$ are the Lagrange multipliers obtained from the \ac{pccd} $\Lambda$ equations, $\hat H$ represents the molecular Hamiltonian{, and we explicitly use $\alpha$- and $\beta$-spin labels to stress the pair nature of the ansatz}. 
The bra state $\langle \Phi_{i \bar{i}}^{a \bar{a}}|$ corresponds to a pair-excited Slater determinant defined as $| \Phi_{i \bar{i}}^{a \bar{a}}\rangle = \hat{a}_a^\dagger \hat{a}_{\bar{a}}^\dagger \hat{a}_{\bar{i}} \hat a_i | \Phi_0 \rangle$.~\cite{tamar-pccd-jcp-2014,ps2-ap1rog-jcp-2014,ap1rog-non-variational-orbital-optimization-jctc-2014}
The resulting \ac{pccd}-optimized natural orbitals (that is, those with a vanishing orbital gradient) are strongly localized. 
We should stress that in this work we consider  \ac{pccd} mainly as a means to provide localized orbitals for both the occupied and virtual subspaces, respectively.

%%%%%%%%%%%%%%%%%%%%%%%%%%%%%%%%%%%%%%%%%%%%%%%%%%%%%%%%%%%%%%%%%%%%
%%%%%%%%%%%%%%%%%%%%%%%%%%%%%%%%%%%%%%%%%%%%%%%%%%%%%%%%%%%%%%%%%%%%
%%%%%%%%%%%%%%%%%%%%%%%%%%%%%%%%%%%%%%%%%%%%%%%%%%%%%%%%%%%%%%%%%%%%
% \section{Results and Discussion}\label{sec:results}
%%%%%%%%%%%%%%%%%%%%%%%%%%%%%%%%%%%%%%%%%%%%%%%%%%%%%%%%%%%%%%%%%%%%
%%%%%%%%%%%%%%%%%%%%%%%%%%%%%%%%%%%%%%%%%%%%%%%%%%%%%%%%%%%%%%%%%%%%
%%%%%%%%%%%%%%%%%%%%%%%%%%%%%%%%%%%%%%%%%%%%%%%%%%%%%%%%%%%%%%%%%%%%

%%%%%%%%%%%%%%%%%%%%%%%%%%%%%%%%%%%%%%%%%%%%%%%%%%%%%%%%%%%%%%%%%%%%
%%%%%%%%%%%%%%%%%%%%%%%%%%%%%%%%%%%%%%%%%%%%%%%%%%%%%%%%%%%%%%%%%%%%
% \subsection{Intermolecular \ac{ct}}
%%%%%%%%%%%%%%%%%%%%%%%%%%%%%%%%%%%%%%%%%%%%%%%%%%%%%%%%%%%%%%%%%%%%
%%%%%%%%%%%%%%%%%%%%%%%%%%%%%%%%%%%%%%%%%%%%%%%%%%%%%%%%%%%%%%%%%%%%

\begin{figure}[!ht]
\centering
\includegraphics[width=\textwidth]{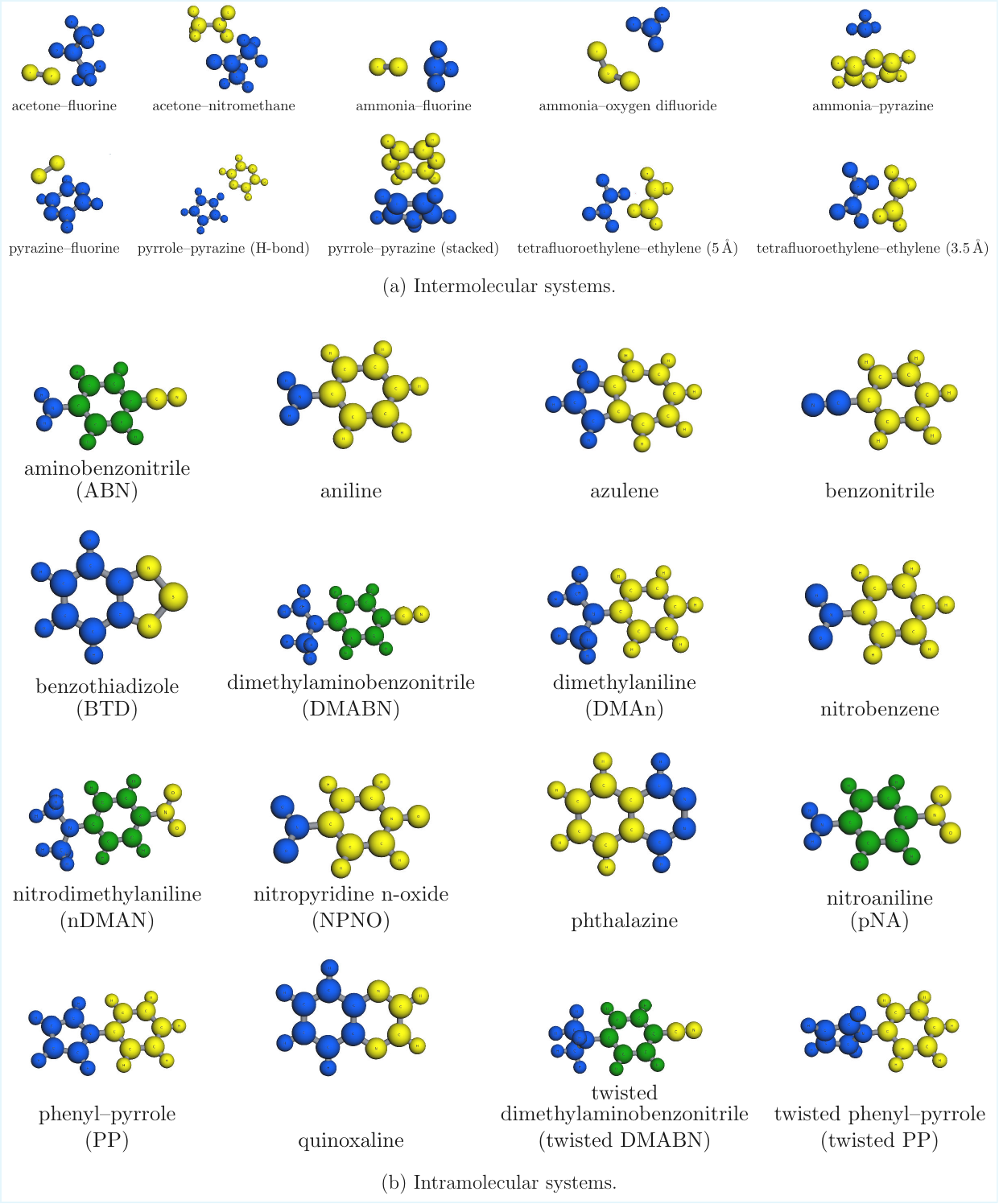}
\caption{All investigated systems exhibiting \ac{ct}. 
Colours indicate molecular roles: blue for the donor, green for the bridge, and yellow for the acceptor domain.
The visualisations were generated using 3Dmol via \ac{daisy}’s graphical user interface.}
\label{fig:2}
\end{figure}

We first assess the \ac{ct} analysis based on our automated domain-based framework discussed above for a set of 10 two-component molecular systems presented in Figure~\ref{fig:2}(a). 
Table~\ref{tab:1} shows the \ac{daisy} \ac{ct} analysis results for \ac{eom}-\ac{ccsd} compared to \ac{ccsd} reference values from a previous study.~\cite{ct-benchmark-kozma-jctc-2020}
{The corresponding comparison using \mbox{\ac{pccds}} is provided in the SI.}
While our \ac{ct} character is reported as the one-directional \ac{ct} between the donor and acceptor fragments, i.e., the non-local component of the charge redistribution or the CT matrix element for the A$\to$B transition,
and displayed in fractional values, the previous study defines their $\omega_{CT}$ as the weight of configurations with charges separated on different fragments. 
In both cases, the excitation character is more local when the corresponding value is close to 0, whereas values approaching 1 indicate a stronger \ac{ct} character.

%%%%%%%%%%%%%%%%%%%%%%%%%%%%%%%%%%%%%%%%%%%%%%%%%%%%%%%%%%%%%%%%%%%%
%%%%%%%%%%%%%%%%%%%%%%%%%%%%%%%%%%%%%%%%%%%%%%%%%%%%%%%%%%%%%%%%%%%%

%%%%%%%%%%%%%%%%%%%%%%%%%%%%%%%%%%%%%%%%%%%%%%%%%%%%%%%%%%%%%%%%%%%%
%%%%%%%%%%%%%%%%%%%%%%%%%%%%%%%%%%%%%%%%%%%%%%%%%%%%%%%%%%%%%%%%%%%%
\begin{table*}[!ht]
\centering
\small
\begin{tabular}{lccccc}
\hline
 &  & \multicolumn{2}{c}{CCSD(HF)~\cite{ct-benchmark-kozma-jctc-2020}} & \multicolumn{2}{c}{CCSD(pCCD)} \\
\hline
Molecule & Basis set & $EE$ & $\omega_{CT}$ & $EE$ & CT \\
\hline

acetone--fluorine & DZ & 6.28 & 0.96 & 6.33 & 0.94 \\
 & TZ &  &  & 6.02 & 0.93 \\
\hline

acetone--nitromethane & DZ & 6.75 & 0.80 & 6.84 & 0.69 \\
 & TZ &  &  & 6.64 & 0.68 \\
\hline

ammonia--fluorine & DZ & 6.90 & 0.76 & 6.87 & 0.79 \\
 & TZ &  &  & 6.63 & 0.79 \\
\hline

ammonia--oxygendifluoride & DZ & 7.33 & 0.86 & 7.37 & 0.84 \\
 & TZ &  &  & 7.33 & 0.61 \\
\hline

ammonia--pyrazine & DZ & 7.93 & 0.63 & 7.95 & 0.61 \\
 & TZ &  &  & 7.81 & 0.75 \\
\hline

pyrazine--fluorine & DZ & 6.73 & 0.64 & 6.74 & 0.64 \\
 &  & 6.77 & 0.98 & 6.77 & 0.94 \\
 & TZ &  &  & 6.28 & 0.90 \\
 &  &  &  & 6.37 & 0.95 \\
\hline

pyrrole--pyrazine (H-bond) & DZ & 5.60 & 1.00 & 5.64 & 0.96 \\
 &  & 6.32 & 0.97 & 6.32 & 0.92 \\
 &  & 6.47 & 0.99 & 6.51 & 0.95 \\
\hline

pyrrole--pyrazine (stacked) & DZ & 5.68 & 0.84 & 5.69 & 0.81 \\
 &  & 6.22 & 0.66 & 6.24 & 0.62 \\
 &  & 6.52 & 0.61 & 6.51 & 0.58 \\
\hline

tetrafluoroethylene--ethylene (5\,\AA) & DZ & 10.87 & 0.99 & 10.87 & 0.97 \\
 & TZ &  &  & 10.54 & 0.94 \\
\hline

tetrafluoroethylene--ethylene (3.5\,\AA) & DZ & 9.05 & 0.34 & 9.08 & 0.36 \\
 & TZ &  &  & 8.66 & 0.35 \\
\hline

\end{tabular}
\caption{
    Intermolecular \ac{ct} character (all \ac{ct} values are of 1$\rightarrow$2 character, i.e., from the first fragment listed in the molecular name to the second) and excitation energies ($EE$ in eV) calculated with cc-pVDZ (DZ) and cc-pVTZ (TZ) basis sets using \ac{eom}-\ac{ccsd}(\ac{pccd}) and reference EOM/LR-\ac{ccsd}(HF) data.
    Missing values have not been computed.
    (HF): indicates that canonical Hartree--Fock orbitals are used to construct the reference determinant.
    (pCCD): indicates a reference determinant constructed with pCCD-optimized orbitals.
    {The corresponding comparison with \mbox{\ac{pccds}} data is available in Table S4 the SI.}
}
\label{tab:1}
\end{table*}
%%%%%%%%%%%%%%%%%%%%%%%%%%%%%%%%%%%%%%%%%%%%%%%%%%%%%%%%%%%%%%%%%%%%

For the intermolecular \ac{ct} analysis, Table~\ref{tab:1} reports the values obtained using the hard partitioning scheme, while a weighted \ac{ct}$_w$ analysis is provided in the SI for the cc-pVTZ basis set. 
Since the systems considered here consist of two separated molecular fragments, both partitioning schemes yield very similar results. 
For this reason, the comparison is limited to the cc-pVTZ basis set, and only the hard \ac{ct} values are discussed in the main text.
The \ac{eom}-\ac{ccsd} states calculated using \ac{pccd} orbitals (labeled as \ac{eom}-\ac{ccsd}(\ac{pccd})), identified by \ac{daisy} as \ac{ct} states, show excellent agreement with the reference EOM/LR-CCSD results computed using canonical \ac{hf} orbitals (EOM/LR-CCSD(\ac{hf})).
{These findings confirm that the \mbox{\ac{ct}} characterization is not tied to a particular orbital representation of the \mbox{\ac{eom}}-\mbox{\ac{ccsd}} states. 
However, the analysis framework itself presupposes a localized orbital basis, for which the \mbox{\ac{pccd}}‑optimized orbitals provide a consistent and reliable choice.}
The excitation energy differences between \ac{eom}-\ac{ccsd}(\ac{pccd}) and EOM/LR-CCSD(\ac{hf}) results are typically within chemical accuracy (around 0.04 eV; a larger error of 0.1 eV was found for acetone--nitromethane), confirming that \ac{eom}-\ac{ccsd} is largely independent of the choice of orbitals in the reference wavefunction.

As expected, the excitation energies obtained with \ac{pccds} {(see the SI)} are generally higher than their \ac{eom}-\ac{ccsd} equivalents. The differences between them are around 2 and 6 eV. Overall, the \ac{ct} values predicted with \ac{pccds} are systematically lower than those obtained with \ac{eom}-\ac{ccsd} and do not quantitatively coincide. 
However, the underlying \ac{ct} character, such as the direction and extent of \ac{ct}, is qualitatively preserved.
Increasing the basis set size from DZ to TZ-quality, overall yields similar \ac{ct} weights (exceptions can be found for ammonia--oxygendifluoride, pyrazine--fluorine, and pyrazine--fluorine).
This demonstrates that the \ac{daisy}-based \ac{ct} analysis is more basis-set {insensitive} (compared to excitation energies) and that \ac{pccds} provides a cheap, but reliable alternative to \ac{eom}-\ac{ccsd} {for extracting CT indices}, offering comparable qualitative performance at a significantly lower computational cost.

%%%%%%%%%%%%%%%%%%%%%%%%%%%%%%%%%%%%%%%%%%%%%%%%%%%%%%%%%%%%%%%%%%%%
%%%%%%%%%%%%%%%%%%%%%%%%%%%%%%%%%%%%%%%%%%%%%%%%%%%%%%%%%%%%%%%%%%%%
% \subsection{Intramolecular \ac{ct}}
%%%%%%%%%%%%%%%%%%%%%%%%%%%%%%%%%%%%%%%%%%%%%%%%%%%%%%%%%%%%%%%%%%%%
%%%%%%%%%%%%%%%%%%%%%%%%%%%%%%%%%%%%%%%%%%%%%%%%%%%%%%%%%%%%%%%%%%%%

\begin{figure}[!ht]
\centering
\includegraphics[width=0.9\textwidth]{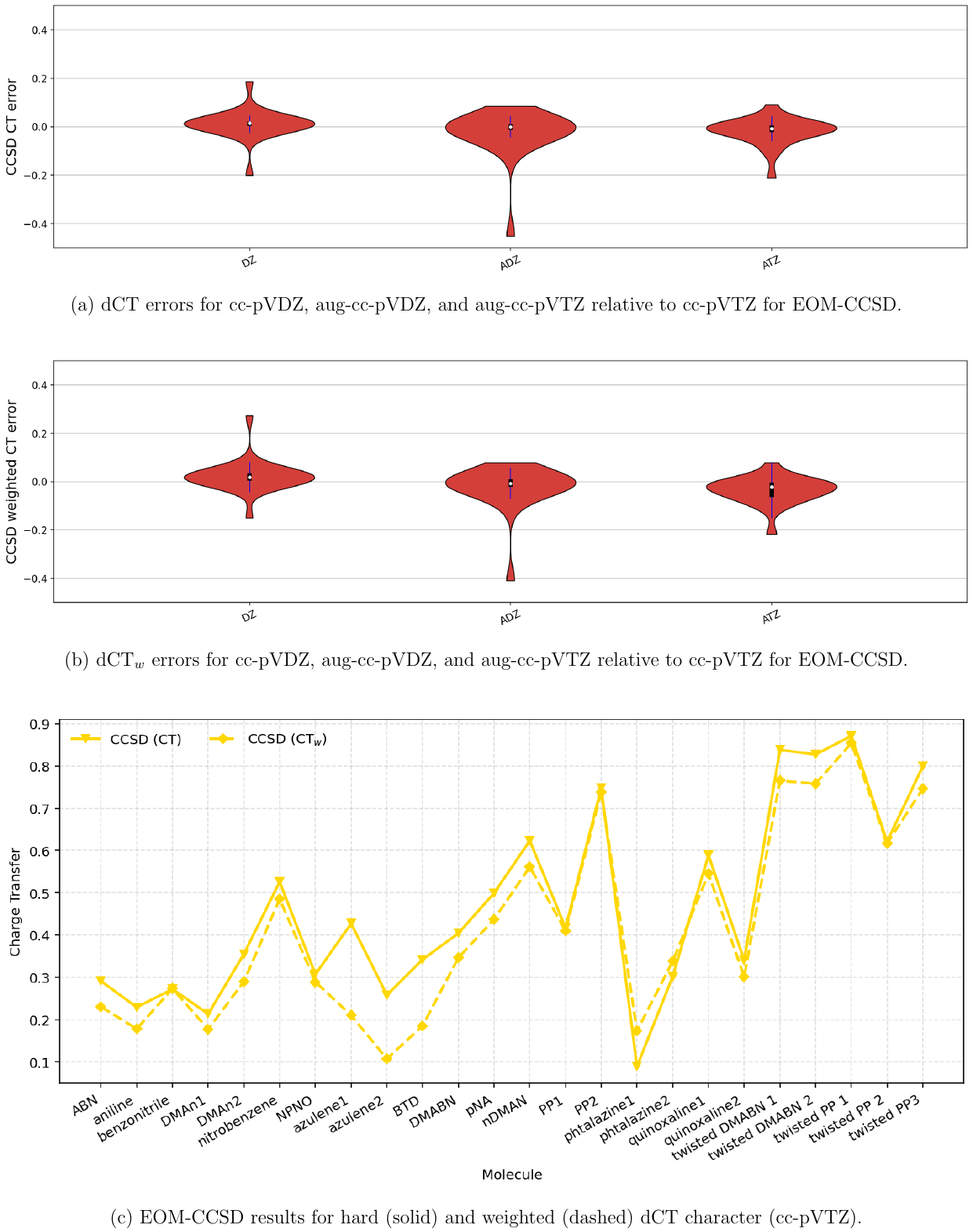}
\caption{\ac{ct} analysis of excited states for \ac{eom}-\ac{ccsd}. The upper panels show violin plots of dCT and dCT$_w$ errors for different basis sets relative to cc-pVTZ. The lower panel compares \ac{eom}-\ac{ccsd} and \ac{pccds} results for strict and weighted \ac{dct} character.
{For systems with multiple \mbox{\ac{ct}} states, numerical labels denote the first, second, and subsequent states, ordered by excitation energy.}
}
\label{fig:3}
\end{figure}

As a second test set, we benchmark our \ac{daisy}-driven {flexible,} automated domain-based \ac{ct} framework for the intramolecular case comprising 16 molecules shown in Figure~\ref{fig:2}(b). %to investigate their intramolecular \ac{ct} character. 
In this particular case, the investigated systems were divided into donor and acceptor domains, and in some cases, a bridge domain was included between them.
The \ac{ct} states were selected based on reference data from a previous study.~\cite{intramolecular-ct-loos-jctc-2021} 
Our results obtained with \ac{eom}-\ac{ccsd} using the cc-pVTZ basis set are summarized in Table~\ref{tab:2} and Figure~\ref{fig:3}(c), while the basis set dependence for \ac{eom}-\ac{ccsd} (for both hard \ac{dct} and weighted \ac{dct}$_w$) is illustrated in Figures~\ref{fig:3}(a) and (b).
The corresponding analysis for \ac{pccds}, including both basis set effects and deviations from \ac{eom}-\ac{ccsd}, is shown in {Table S4 and} Figure S2 in the SI.
We should stress that, for all methods and basis sets, we consider the \ac{dct} values only (see eq.~\eqref{eq:dct}), which quantifies the net charge flow from the donor to the bridge (if present) to the acceptor domain.

Overall, the {flexible,} automated domain-based CT framework predicts a wide range of CT states, from weak charge-transfer character (e.g., aniline, indicated by a ``+'' in the Table) to strongly charge-separated states (e.g., nitrobenzene, pNA, or nDMAN, indicated by a ``+++'' in the Table) to almost pure \ac{ct} states (e.g., PP and twisted PP and DMABN systems, indicated by a ``++++'' in the Table).
As expected, twisted systems exhibit a strong charge separation and feature excited states of almost pure \ac{dct} character.
Slightly smaller, albeit still significant \ac{dct} values are obtained for donor--bridge--acceptor assemblies, such as pNA or nDMAN.
The weakest \ac{dct} values are observed for 2-domain molecules comprising a donor--ring configuration, followed by undoped, C-only systems (like azulene).

Furthermore, both the hard and weighted CT values are rather basis-set {insensitive} (see Figure~\ref{fig:3}), as the error distributions for both descriptors are closely centered around zero for all basis sets considered (with a standard deviation at around 3\% for larger basis sets). 
Specifically, for the hard \ac{ct} flavor, the spread of the distribution remains relatively uniform across the cc-pVDZ, aug-cc-pVDZ, and aug-cc-pVTZ basis set, indicating a stable behavior with respect to basis set enlargement.
In contrast, the weighted $\mathrm{CT}_w$ exhibits slightly broader distributions for smaller basis sets, particularly for aug-cc-pVDZ, where a few outliers are visible. 
Nevertheless, these deviations remain modest and do not affect the overall conclusion that basis set effects are minor.
Importantly, the absence of larger shifts in the median values confirms that no significant bias is introduced by the choice of basis set. 
This robustness is essential for practical applications, as it allows the use of smaller basis sets without compromising the qualitative interpretation of the \ac{ct} character.

%%%%%%%%%%%%%%%%%%%%%%%%%%%%%%%%%%%%%%%%%%%%%%%%%%%%%%%%%%%%%%%%%%%%
\begin{table}[ht!]
\centering
\begin{tabular}{lcccc}
\hline
molecule & dCT char. & $EE$ & dCT & dCT$_w$ \\
\hline
ABN             & ++    & 5.41 & 0.29 & 0.23 \\
aniline         & ++    & 5.99 & 0.23 & 0.18 \\
benzonitrile    & ++    & 7.33 & 0.27 & 0.27 \\
DMAn 1          & ++    & 4.66 & 0.21 & 0.18 \\
DMAn 2          & ++    & 5.68 & 0.36 & 0.29 \\
nitrobenzene    & +++   & 5.77 & 0.53 & 0.48 \\
NPNO            & ++    & 4.45 & 0.31 & 0.29 \\
azulene 1       & ++    & 4.03 & 0.41 & 0.20 \\
azulene 2       & +     & 4.83 & 0.24 & 0.10 \\
BTD             & ++    & 4.67 & 0.34 & 0.19 \\
DMABN           & ++    & 5.10 & 0.40 & 0.35 \\
pNA             & +++   & 4.81 & 0.50 & 0.44 \\
nDMAN           & +++   & 4.53 & 0.62 & 0.56 \\
PP 1            & +++   & 5.87 & 0.42 & 0.41 \\
PP 2            & ++++  & 6.56 & 0.75 & 0.74 \\
phtalazine 1    & ++    & 4.28 & 0.09 & 0.17 \\
phtalazine 2    & ++    & 4.64 & 0.30 & 0.34 \\
quinoxaline 1   & +++   & 5.04 & 0.59 & 0.54 \\
quinoxaline 2   & ++    & 5.98 & 0.34 & 0.30 \\
twisted DMABN 1 & ++++  & 4.42 & 0.84 & 0.77 \\
twisted DMABN 2 & ++++  & 5.19 & 0.83 & 0.76 \\
twisted PP 1    & ++++  & 6.13 & 0.87 & 0.85 \\
twisted PP 2    & +++   & 6.22 & 0.62 & 0.62 \\
twisted PP 3    & ++++  & 6.33 & 0.80 & 0.75 \\
\hline
ME (DZ)      & -- & --   & 0.01 & 0.02 \\
SD (DZ)      & -- & --   & 0.06 & 0.07 \\
ME (ADZ)     & -- & --   & -0.03 & -0.02 \\
SD (ADZ)     & -- & --   & 0.10 & 0.10 \\
ME (ATZ)     & -- & --   & -0.02 & -0.04 \\
SD (ATZ)     & -- & --   & 0.06 & 0.06 \\
\hline
\end{tabular}
\caption{
    Intramolecular hard (dCT) and weighted ($\mathrm{dCT}_w$) \ac{dct} character and excitation energies ($EE$ in eV) calculated with the cc-pVTZ basis set using \ac{eom}-\ac{ccsd}.
    The \ac{dct} is evaluated according to eq.~\eqref{eq:dct}.
    {For systems with multiple \mbox{\ac{ct}} states, numerical labels denote the first, second, and subsequent states, ordered by excitation energy.}
    The qualitative \ac{dct} character (\ac{dct} char.) is assigned based on the weighted \ac{eom}-\ac{ccsd} $\mathrm{dCT}_w$ values: weak (+) for $|\mathrm{dCT}_w |\leq 0.10$, moderate (++) for $0.10 < |\mathrm{dCT}_w | \leq 0.35$, strong (+++) for $0.35 < |\mathrm{dCT}_w | \leq 0.7$, and (mostly) pure (++++) for $0.7 < |\mathrm{dCT}_w | \leq 1.0$. 
    Mean error (ME) and standard deviation (SD) are also provided for the cc-pVDZ (DZ), aug-cc-pVDZ (ADZ), and aug-cc-pVTZ (ATZ) basis sets, using the cc-pVTZ results as reference values. 
    The distribution of these statistics is illustrated by the violin plots shown in Figure~\ref{fig:3}.
    {The parallel analysis for \mbox{\ac{pccds}} data is available in Table S5 and Figure S2 the SI.}  
}
\label{tab:2}
\end{table}
%%%%%%%%%%%%%%%%%%%%%%%%%%%%%%%%%%%%%%%%%%%%%%%%%%%%%%%%%%%%%%%%%%%%
Furthermore, comparing the hard \ac{dct} and weighted $\mathrm{dCT}_w$ measures shown in Figure~\ref{fig:3} reveals a rather consistent description between both domain accumulation recipes across systems of varying size and complexity. However, we recommend the weighted $\mathrm{CT}_w$ framework, especially if the molecules to be partitioned are small and feature domains that are connected by more than one interatomic connection.
In particular, $\mathrm{CT}_w$ is less sensitive to the partitioning into domains and remains additive when multiple charge-transfer pathways are present. 
This becomes particularly evident for systems such as phthalazine, where different domain partitions (e.g., two versus three domains; see also Table S2 in the SI) can be considered. 
In this case, the weighted $\mathrm{CT}_w$ measure provides a consistent and additive description, whereas the hard \ac{ct} definition leads to a more fragmented and less transparent picture.
This can be understood in terms of \ac{ao} contributions to each \ac{mo}.
If two domain \ac{ao}s contribute (almost equally) to a localized \ac{mo}, the weighted $\mathrm{CT}_w$ will distribute the domain contributions of the CI vector component accordingly, while the hard $\mathrm{CT}$ measure will assign it to a single domain only.
We should note, however, that this domain-assignment problem becomes negligible in larger molecules partitioned into a minimal number of domains.

%%%%%%%%%%%%%%%%%%%%%%%%%%%%%%%%%%%%%%%%%%%%%%%%%%%%%%%%%%%%%%%%%%%%
%%%%%%%%%%%%%%%%%%%%%%%%%%%%%%%%%%%%%%%%%%%%%%%%%%%%%%%%%%%%%%%%%%%%

Considering simpler \ac{eom} approximations, the \ac{pccds} results qualitatively reproduce the trends observed at the \ac{eom}-\ac{ccsd} level (see also Figure~{S3}).
However, the magnitude of the \ac{ct} character is systematically underestimated.
Despite this quantitative discrepancy, both methods yield consistently similar qualitative behaviour.
This observation supports the use of \ac{pccds} as a more computationally efficient alternative for qualitative \ac{ct} analysis, focusing on changes in the \ac{ct} character upon, for instance, doping or domain modifications/substitutions.
Finally, the basis set analysis shown in Figure S2 in the SI confirms that both \ac{ct} descriptors are largely insensitive to the choice of basis set. 
The small mean errors and standard deviations indicate that the cc-pVTZ basis provides a reliable reference, and that lower-cost basis sets can be used without significantly affecting the qualitative conclusions.
This is further supported by the violin plots in Figure S2 in the SI, where the distributions of the \ac{pccds} errors remain narrow and largely unchanged across different basis sets. 

%%%%%%%%%%%%%%%%%%%%%%%%%%%%%%%%%%%%%%%%%%%%%%%%%%%%%%%%%%%%%%%%%%%%%%%%%%%%%%
%%%%%%%%%%%%%%%%%%%%%%%%%%%%%%%%%%%%%%%%%%%%%%%%%%%%%%%%%%%%%%%%%%%%%%%%%%%%%%
%%%%%%%%%%%%%%%%%%%%%%%%%%%%%%%%%%%%%%%%%%%%%%%%%%%%%%%%%%%%%%%%%%%%%%%%%%%%%%
% \section{Conclusions and Outlook}\label{sec:conclusions}
%%%%%%%%%%%%%%%%%%%%%%%%%%%%%%%%%%%%%%%%%%%%%%%%%%%%%%%%%%%%%%%%%%%%%%%%%%%%%%
%%%%%%%%%%%%%%%%%%%%%%%%%%%%%%%%%%%%%%%%%%%%%%%%%%%%%%%%%%%%%%%%%%%%%%%%%%%%%%
%%%%%%%%%%%%%%%%%%%%%%%%%%%%%%%%%%%%%%%%%%%%%%%%%%%%%%%%%%%%%%%%%%%%%%%%%%%%%%
In this work, we developed and benchmarked a {flexible,} automated framework for analyzing \ac{ct} in excited states, which is integrated into the \ac{daisy} module of the \ac{pybest} quantum chemistry package.\cite{pybest-paper-cpc-2021,pybest-paper-cpc-2024}
Two complementary domain accumulation schemes are available, namely a strict (hard) partitioning and a weighted scheme designed for smaller molecules and large domain numbers.
We benchmark our approach for \ac{eom}-\ac{ccsd}, various basis set sizes, and \ac{ct} states (intra- and intermolecular ones).
For intermolecular \ac{ct}, both the excitation energies and the \ac{ct} character obtained with \ac{eom}-\ac{ccsd} are largely independent of the chosen reference determinant (or \ac{mo} basis), including localized \ac{pccd}-optimized and canonical \ac{hf} orbitals.
Although the excitation energies computed with \ac{pccds} are generally higher by approximately 2--6 eV, the overall \ac{ct} character remains very similar to \ac{eom}-\ac{ccsd} results.
In most intermolecular cases, the \ac{ct} character exhibits a strong basis-set {insensitivity}, making the proposed automated domain-based CT analysis a suitable tool to efficiently and reliably resolve the excited-state character.
Furthermore, both the hard and weighted domain accumulation schemes yield consistent results for the intermolecular \ac{ct}.
Due to its additivity with respect to domain contributions, the weighted recipe is, however, preferred, in particular for smaller molecules.
Although some variations, particularly for systems such as azulene (\ac{eom}-\ac{ccsd} and \ac{pccds}) and BTD (\ac{eom}-\ac{ccsd}), are observed, the overall trends remain consistent across all investigated CT systems.
Due to the basis-set independence of the intramolecular \ac{ct}, already smaller basis sets can be employed to resolve domain-based excitations.
Finally, in comparison to \ac{eom}-\ac{ccsd}, the \ac{pccds} method generally yields lower \ac{ct} values for intramolecular excitations, while preserving the qualitative trends observed with the former method.
It thus represents a cheap alternative to dissecting excited-state characters into domain contributions, which makes it especially appealing for large-scale systems.
{While \mbox{\ac{pccds}} is more approximate and not intended to replace broadly available approaches such as TD-DFT to predict reliable excitation energies, it offers a computationally economical, internally consistent perspective on CT states and the evolution of the CT character upon structural modification.}
{Diffuse basis functions remain essential for accurately describing various \mbox{\ac{ct}} excitations. 
This challenge is intrinsic to the excited‑state problem rather than a deficiency of the proposed methodology. 
The unique potential of pCCD to strongly localize virtual orbitals alleviates some of the complications introduced by very diffuse functions, although it cannot eliminate them entirely.}
{In future work, we plan to probe the sensitivity of the DAISpY workflow on different choices of localization procedures, which might further accelerate the proposed domain-based CT decomposition, skipping the pCCD-based localization step.}
{Furthermore, we will compare the present scheme with alternative \mbox{\ac{ct}} analysis approaches, including density‑based schemes, once excited‑state densities become available in our workflow. 
Such comparisons will allow us to further assess the robustness and generality of the proposed methodology.}
Overall, the presented framework provides a robust and efficient tool for a systematic and {flexible,} automated, domain-based excited-state analysis.
{Owing to its formulation in terms of CI-type wavefunction amplitudes, the scheme is readily applicable to a broad range of excited-state methods, avoiding explicit left-eigenvector or excited-state density evaluations. 
Finally, the proposed approach can be straightforwardly extended to higher-order excitations or other excited-state models in the future.}

%%%%%%%%%%%%%%%%%%%%%%%%%%%%%%%%%%%%%%%%%%%%%%%%%%%%%%%%%%%%%%%%%%%%
%%%%%%%%%%%%%%%%%%%%%%%%%%%%%%%%%%%%%%%%%%%%%%%%%%%%%%%%%%%%%%%%%%%%
%%%%%%%%%%%%%%%%%%%%%%%%%%%%%%%%%%%%%%%%%%%%%%%%%%%%%%%%%%%%%%%%%%%%
\section*{Computational Details}\label{sec:comput-det}
%%%%%%%%%%%%%%%%%%%%%%%%%%%%%%%%%%%%%%%%%%%%%%%%%%%%%%%%%%%%%%%%%%%%
%%%%%%%%%%%%%%%%%%%%%%%%%%%%%%%%%%%%%%%%%%%%%%%%%%%%%%%%%%%%%%%%%%%%
%%%%%%%%%%%%%%%%%%%%%%%%%%%%%%%%%%%%%%%%%%%%%%%%%%%%%%%%%%%%%%%%%%%%
All calculations were performed using variationally optimized \ac{pccd} orbitals if not stated otherwise.
In all \ac{cc} calculations, a frozen core approximation was applied, keeping 1s for C, N, O, and F, and 1s-2p for S frozen. 
We employed two \ac{eom} variants to study the \ac{ct} character and to obtain the excitation energies in the considered systems, namely \ac{eom}-\ac{ccsd} and \ac{pccds}.
All calculations were performed with the \ac{pybest} v.2.2.0.dev0 software package\cite{pybest-paper-cpc-2021, pybest-paper-cpc-2024}, using GPU-acceleration for \ac{eom}-\ac{ccsd} and some of the larger \ac{pccd} cases.~\cite{pybest-gpu-jctc-2024}
In the case of intramolecular \ac{ct}, the cc-pVDZ, cc-pVTZ,  aug-cc-pVDZ, and aug-cc-pVTZ basis sets were applied.~\cite{dunning-gaussian-basis-set-jcp-1989,aug-cc-pvtz-jcp-1989} 
For intermolecular \ac{ct}, only cc-pVDZ and cc-pVTZ basis sets were used.
In all calculations, Cholesky-decomposed electron repulsion integrals were used with a Cholesky threshold of $10^{-4}$, which is sufficient to obtain excitation energies with sub-chemical accuracy.~\cite{ea-eom-fpccd-jctc-2025}
The intermolecular \ac{ct} test set is shown in Figure~\ref{fig:2}(a), while Figure~\ref{fig:2}(b) summarizes the intramolecular test set.
The molecular structures were taken from the literature.~\cite{ct-benchmark-kozma-jctc-2020, intramolecular-ct-loos-jctc-2021}
\ac{daisy} as a PyBEST interface was used for the \ac{ct} analysis.~\cite{jj-daispy-inprep-2026}

%%%%%%%%%%%%%%%%%%%%%%%%%%%%%%%%%%%%%%%%%%%%%%%%%%%%%%%%%%%%%%%%%%%%%
%% The same is true for Supporting Information, which should use the
%% suppinfo environment.
%%%%%%%%%%%%%%%%%%%%%%%%%%%%%%%%%%%%%%%%%%%%%%%%%%%%%%%%%%%%%%%%%%%%%
\begin{suppinfo}
The following files are available free of charge:
\begin{itemize}
  \item si.pdf: Weighted vs hard intermolecular directed CT, additivity of the weighted CT framework, violin plots for EOM-pCCD+S, {inter- and intramolecular analysis for EOM-pCCD+S}

  \item si.xlsx: \ac{daisy} data (excitation energies, hard and weighted \ac{ct} values)

\end{itemize}
\end{suppinfo}

%%%%%%%%%%%%%%%%%%%%%%%%%%%%%%%%%%%%%%%%%%%%%%%%%%%%%%%%%%%%%%
\begin{acknowledgement}
%%%%%%%%%%%%%%%%%%%%%%%%%%%%%%%%%%%%%%%%%%%%%%%%%%%%%%%%%%%%%%
M.G.~acknowledges financial support from the SONATA research grant from the National Science Centre, Poland (Grant No. 2023/51/D/ST4/02796). 
Funded/Co-funded by the European Union (ERC, DRESSED-pCCD, 101077420).
Views and opinions expressed are, however, those of the author(s) only and do not necessarily reflect those of the European Union or the European Research Council. Neither the European Union nor the granting authority can be held responsible for them.
We gratefully acknowledge Polish high-performance computing infrastructure PLGrid (HPC Centers: ACK Cyfronet AGH and WCSS) for providing computer facilities and support within computational grant no. PLG/2025/018840.
We thank Iulia Brumboiu for many fruitful discussions.
\end{acknowledgement}
%%%%%%%%%%%%%%%%%%%%%%%%%%%%%%%%%%%%%%%%%%%%%%%%%%%%%%%%%%%%%%

%%%%%%%%%%%%%%%%%%%%%%%%%%%%%%%%%%%%%%%%%%%%%%%%%%%%%%%%%%%%%%%%%%%%%
%%%%%%%%%%%%%%%%%%%%%%%%%%%%%%%%%%%%%%%%%%%%%%%%%%%%%%%%%%%%%%%%%%%%%
\section*{Data Availability Statements}
The data underlying this study are available in the published article and its Supporting Information.
The released version of the PyBEST code is available on Zenodo at \url{https://zenodo.org/records/10069179} and on PyPI at \url{https://pypi.org/project/pybest/}.
A developer, stand-alone version of \ac{daisy} is available on GitLab at
\url{https://gitlab.com/pybest-edev/ct-analysis}.

\section*{Conflicts of Interest}
There are no conflicts to declare.

%%%%%%%%%%%%%%%%%%%%%%%%%%%%%%%%%%%%%%%%%%%%%%%%%%%%%%%%%%%%%%%%%%%%%
\bibliography{p}

\end{document}